# VHDL Implementation of different Turbo Encoder using Log-MAP Decoder

Akash Kumar Gupta and Sanjeet Kumar

**Abstract**— Turbo code is a great achievement in the field of communication system. It can be created by connecting a turbo encoder and a decoder serially. A Turbo encoder is build with parallel concatenation of two simple convolutional codes. By varying the number of memory element (encoder configuration), code rate (1/2 or 1/3), block size of data and iteration, we can achieve better BER performance. Turbo code also consists of interleaver unit and its BER performance also depends on interleaver size. Turbo Decoder can be implemented using different algorithm, but Log –MAP decoding algorithm is less computationaly complex with respect to MAP (maximux a posteriori) algorithm, without compromising its BER performance, nearer to Shannon limit. A register transfer level (RTL) turbo encoder is designed and simulated using VHDL (Very high speed integrated circuit Hardware Description Language). In this paper VHDL model of different turbo encoder are implemented using Log MAP decoder and its performance are compared and verified with corresponding MATLAB simulated results.

## 1 Introduction

Now a days, turbo-codes [1] have received considerable attention and first came in existence in 1993. The turbo encoder consists of two recursive systematic convolutional (RSC) encoders operating in parallel manner with the input data bits. We can design either a 4-state or 8-state turbo encoder, with different rate. An interleaver is used between two systematic convolution encoder as shown in Figure1. Here, we can achieve the rate 1/3, without puncturing and 1/2, with puncturing method. Other code rates are also obtained by puncturing mechanism.

When turbo encoded data is applied to turbo decoder through Base-Band transmission over AWGN channel, Log-Map decoding structure provides performance nearer to Shannon's limit [2] with less complexity [3] with respect SOVA. Due to its powerful error correcting capability, reasonable complexity, and flexibility in terms of different block lengths and code rates, number of memory element etc., turbo code are widely used in different communication system like satellite communication, CDMA, mobile communication etc.

For implementing the turbo decoder, the very first consideration is to select a SISO algorithm, which can give efficient performance. Here, we have chosen Log-Map decoding algorithm [4], which provides good performance with less complexity.

In this paper, we first discussed about the turbo encoder and its different configuration, which we have implemented using BHDL as well as MATLAB simulation.the variation on its configuration with the implementation issue using VHDL [5].

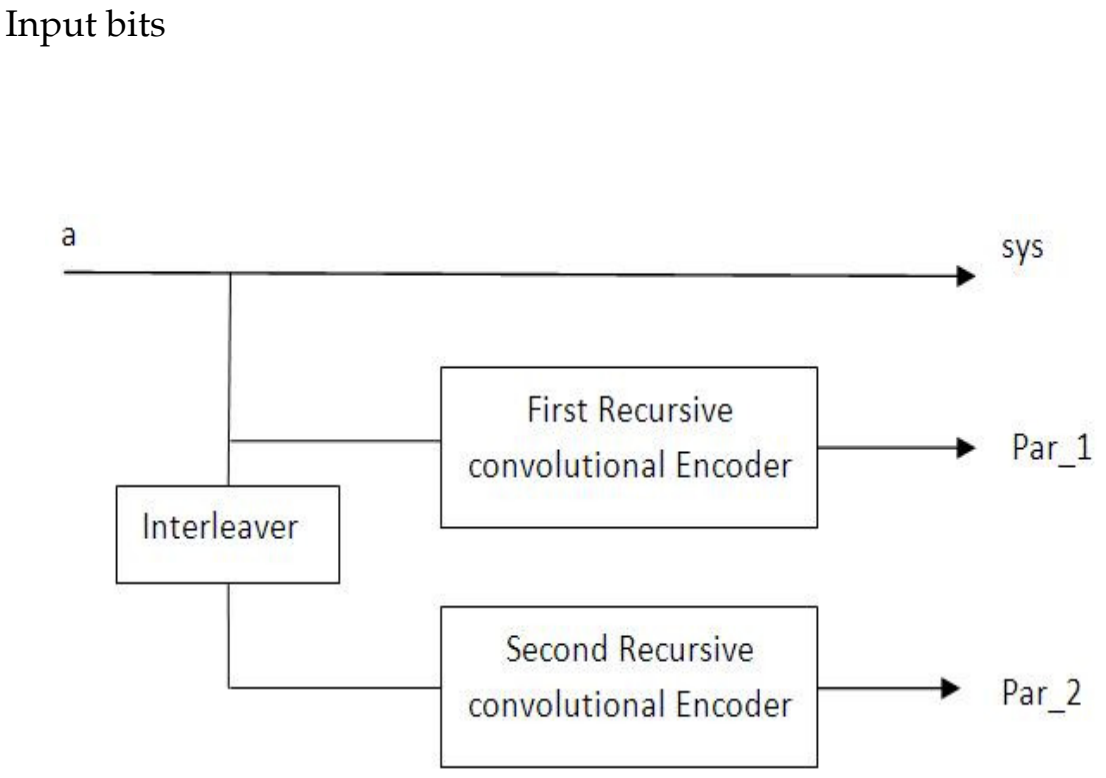

Fig. 1: Turbo encoder without puncturing (Rate 1/3)

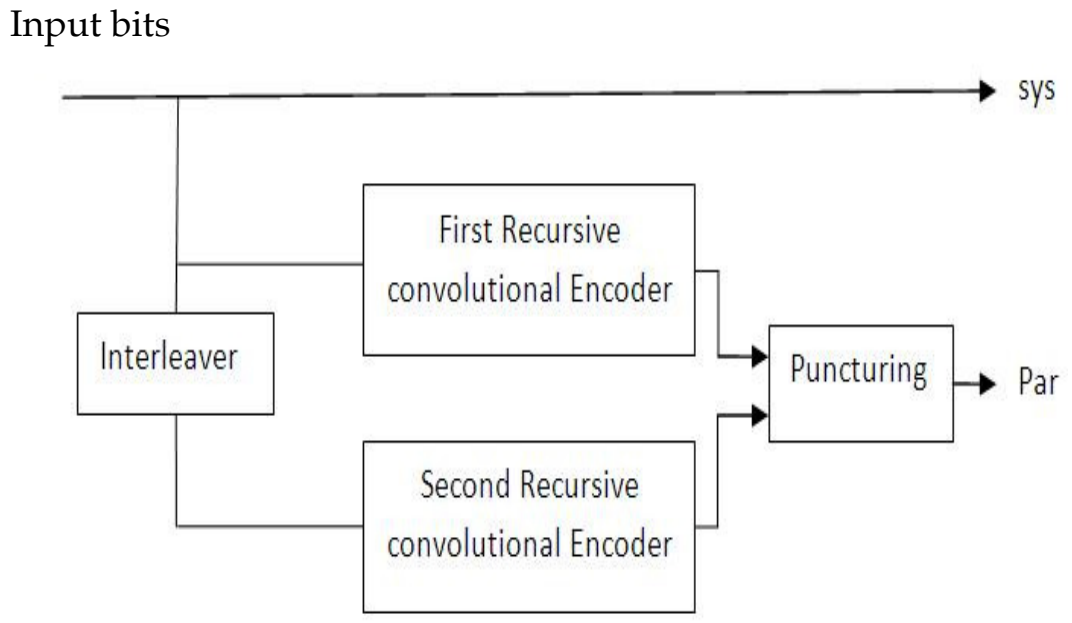

Fig. 2: Turbo encoder with puncturing (Rate 1/2)

Secondary, we discussed about Log-MAP decoder (SISO) algorithm, which reduces the complexity of the system.

Later, we will show the comparison and verification of

- *Akash Kumar Gupta is with the Department of Electronics and Communication Birla Institute of Technology, Mesra, Ranchi, India.*
- *Sanjeet Kumar is with the Department of Electronics and Communication, Birla Institute of technology, Mesra, Ranchi, India.*

the VHDL implemented model with the MATLAB implemented model of system.

## 2 Implementation of turbo Code

A turbo code actually consists of two systematic convolutional encoders [6], which consist of an interleaver unit with a specified encoding structure.

We send n number of input data bits to the encoder. This data goes to the first convolutional encoder whereas an interleaved data is passed through the second convolutional encoder. From Fig. 1, it is clear that systematic bit is same as input data bits and with the help of two RSC encoders, we get the two n bits parity sequences. These three sequences create rate 1/3. The two parity sequences get punctured alternatively and we get the rate 1/2.

Turbo encoded output data bits are passed through AWGN (Additive white Gaussian noise) channel. Due to noise some of data bits may get corrupted and error maybe introduced in the system. These data bits are transferred to the decoder unit where the error is removed and original data is recovered.

### 2.1 Turbo Encoder

We have implemented different configurations of turbo encoder using VHDL.

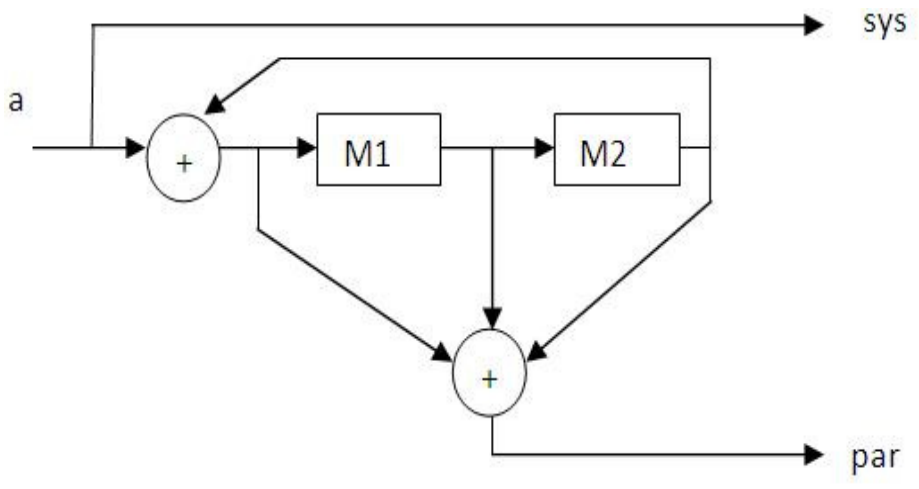

Fig. 3(a) configuration (5, 7)

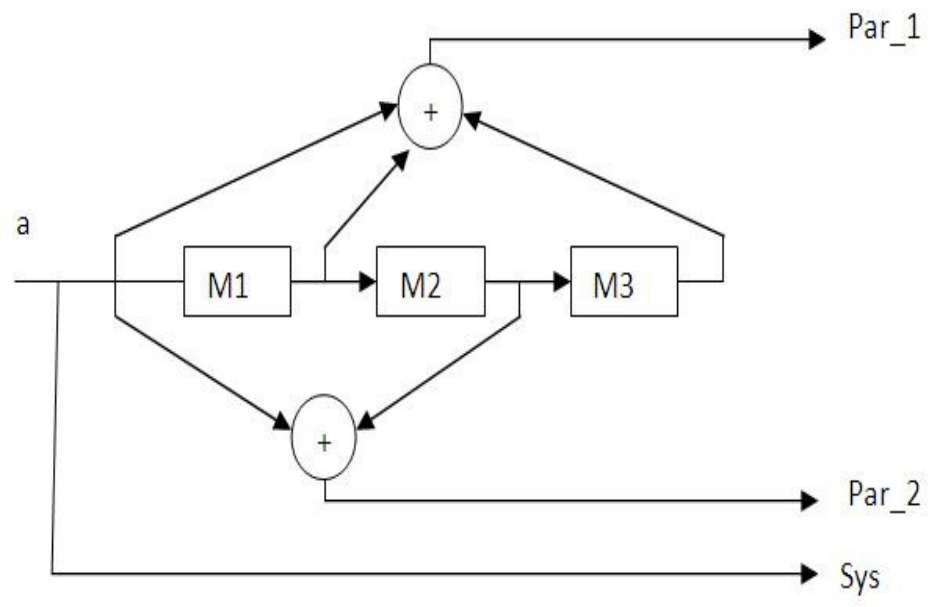

Fig. 3(b) configuration (15, 12, 0)

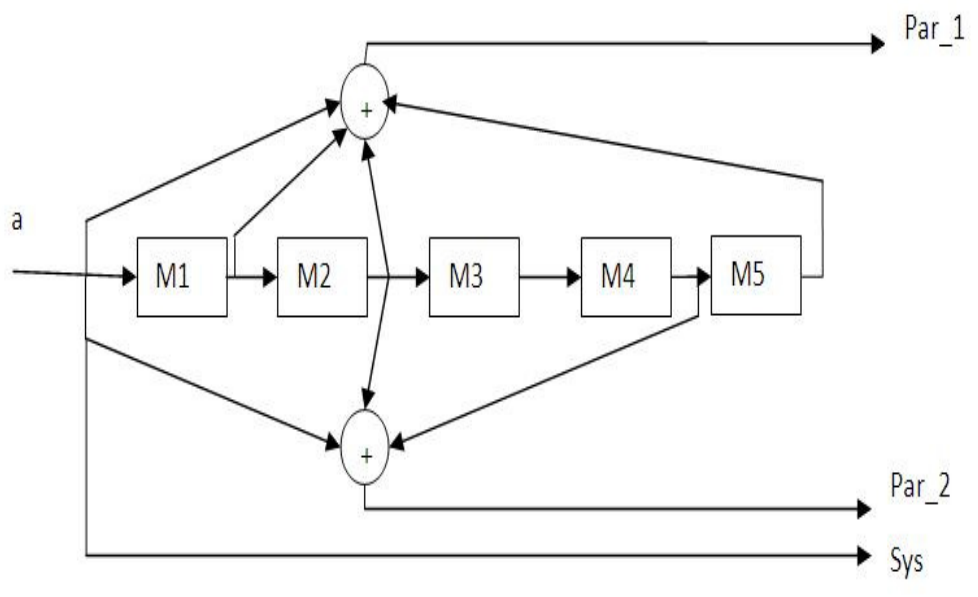

Fig. 3(c) configuration (71, 52, 0)

Performance is highly depending upon memory element. By varying the number of memory elements, we can get different configuration. There is some of the following encoder structure that we have implemented using VHDL as shown in Fig. 3(a), 3(b) and 3(c).

These configurations are used in the forms of binary polynomial, respectively, the first is (101, 111), second (1111, 1010) and third is (111001, 101010).

From the above structures, we get the systematic, par_1 and par_2 (parity) sequences. These are available according to rate 1/3. After puncturing of par_1 and par_2, we get alternatively combined parity sequence, corresponds to rate 1/2.

After implementing the turbo encoder in VHDL, we got its output waveform, shown in fig. 4 and output data bits stored in binary form. At the decoder side, these data bit sequences are used for decoding.

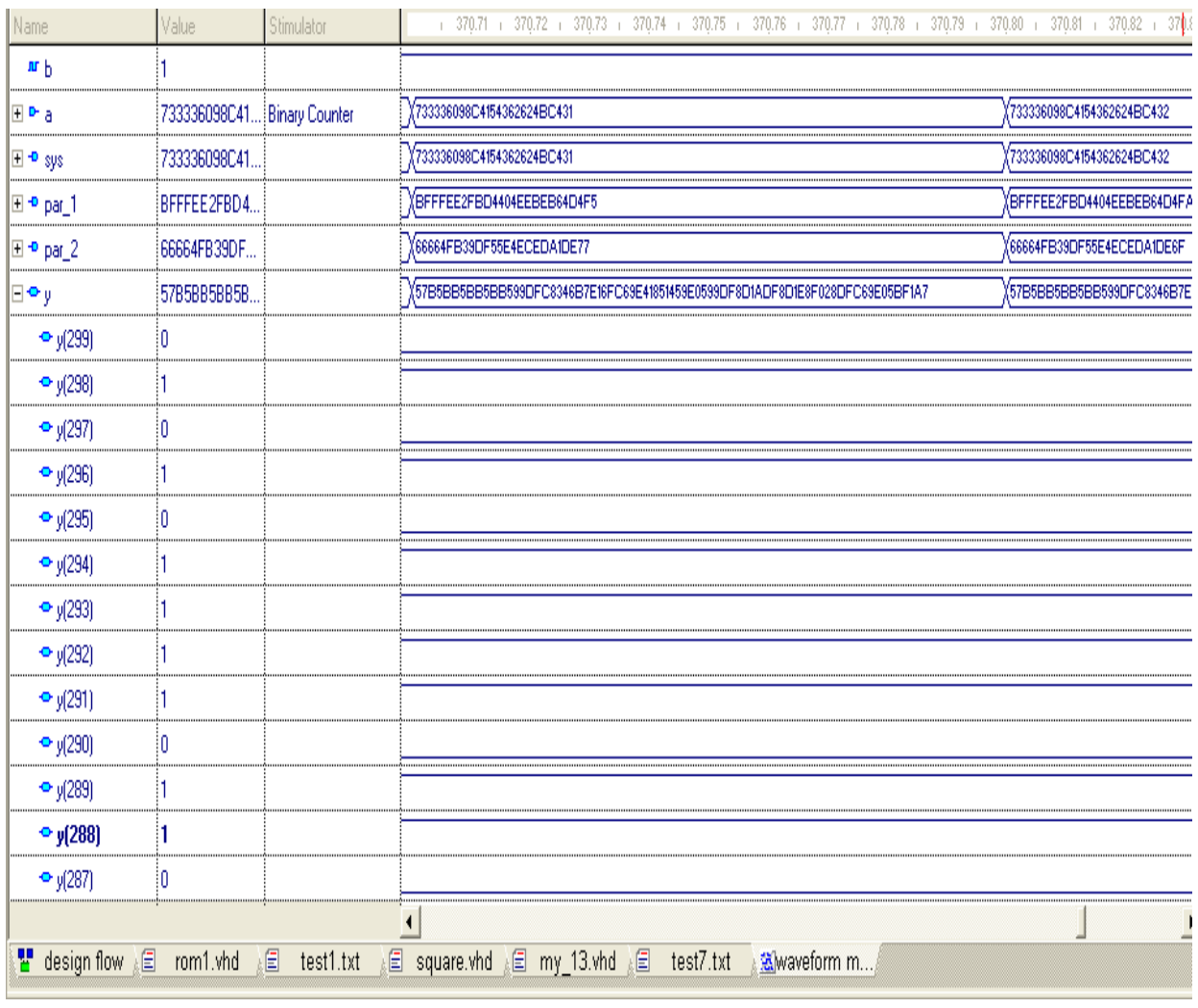

Fig. 4: output waveform for encoder configuration (15, 12, 0)

To get the desired BER performance, we have implemented different configuration of turbo encoder using different interleaver size and analyse its performance.

### 2.2 Interleaver

In turbo code, interleaver unit is a random block that is used to rearrange the input data bits with no repetition.

Interleaver unit is used in both encoder and decoder part. At the encoder side it generates a long block of data, whereas in decoder part it correlates the two SISO decoder and helps to correct the error.

At the decoder side after passing the encoded data from first dcoder some of the errors may get corrected, then we again interleav, this first decoded data and pass through the second decoder. Here, remaining error may get correct. Like this, we are repeating the process for more number of times.

### 2.3 Turbo Decoder

Decoding algorithm can be designed [7] by either an A posteriori probability (APP) method or maximum likelihood method. Turbo decoder consists of two SISO decoders separated by interleaver and de-interleaver, as shown in Fig. 5. Due to noise, encoded output data bit may get corrupted and reach the decoder input as r0 for systematic bit, r1 for parity_1 and r2 for parity_2 respectively.

These inputs are fed to first SISO decoder. First SISO decoder takes as input, the received information data bits sequence r0 and received parity sequence r1, which is generated by RSC encoder 1. The output data seq. from the first SISO decoder is interleaved and goes as an input to second SISO decoder. The second SISO decoder takes as input, r0 as received information bit sequence and r2 as received parity sequence. From the second SISO decoder we get the final output data. To improve the performance of turbo code its output is de-interleaved and sent to first SISO decoder as an input. Then we go for the process repetation. This whole process is also known as iterative decoding [8] process.

After certain iteration the output of decoder stops further performance improvement. There are different decoding algorithms available, here we used Log-MAP decoding algorithm, which uses iterative decoding process.

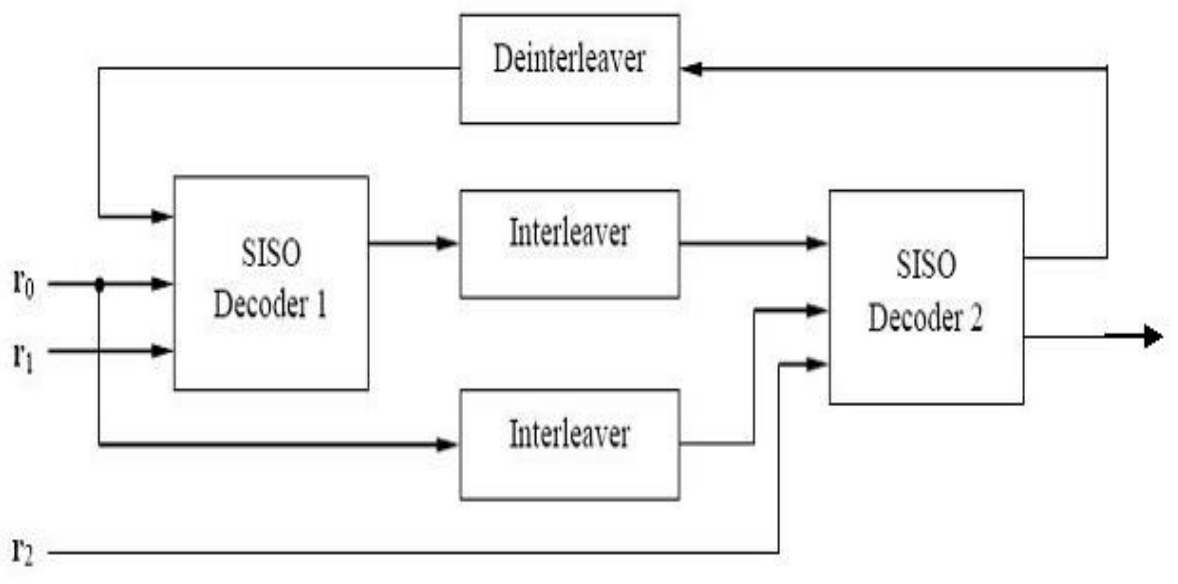

Fig. 5: General diagram of turbo Log MAP decoder

Finally, we take hard decision, to retrieve original data bits.

Now, we calculate the BER (bit error rate) value [8, 10] for different set Eb/No ratio and performance is compared.

## 3 Results and Discussion

We have developed VHDL model of different turbo encoder with Log-MAP decoder. Fig. 5 is showing our MATLAB and VHDL simulated results for (7, 5) configuration of turbo encoder with Rate 1/3 and 8 iteration. This result is compared with result given by L. Hanzo et al. [11], with same configuration. We finding that our simulated VHDL and MATLAB both result are matching with reported result.

To verify VHDL simulation maodel, again we have another encoder configuration shown in Fig. 6. This result shows that for Rate 1/3 our performance is improves by 0.25 dB and shows the performance improvement from rate 1/2 to 1/3. The improvement in rate 1/3 encoder with respect to 1/2 encoder is expected.

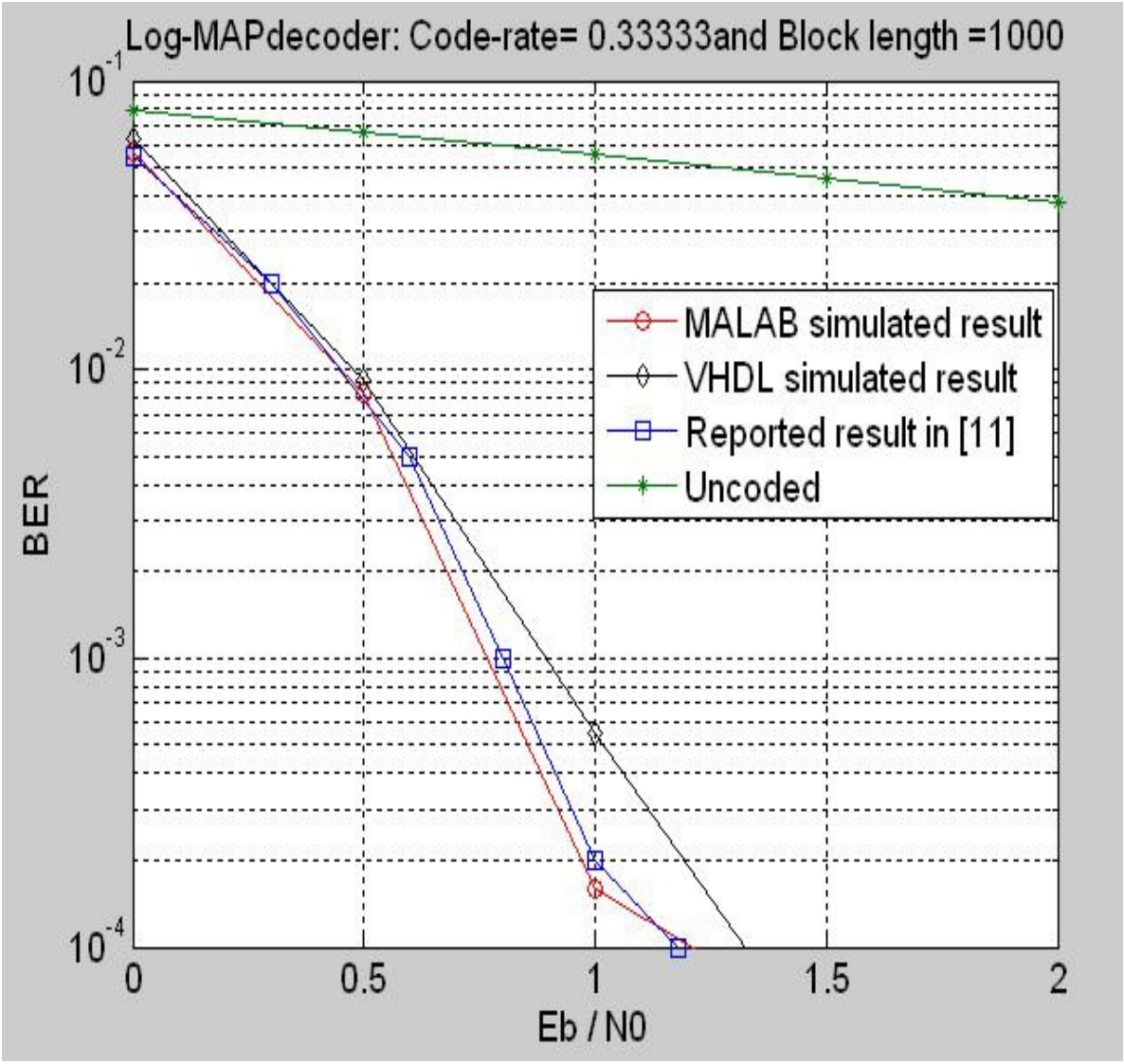

Fig. 5: Comparison of result obtained from MATLAB and VHDL Simulation for the configuration (7, 5) with reported result

This result further verifies our VHDL model.

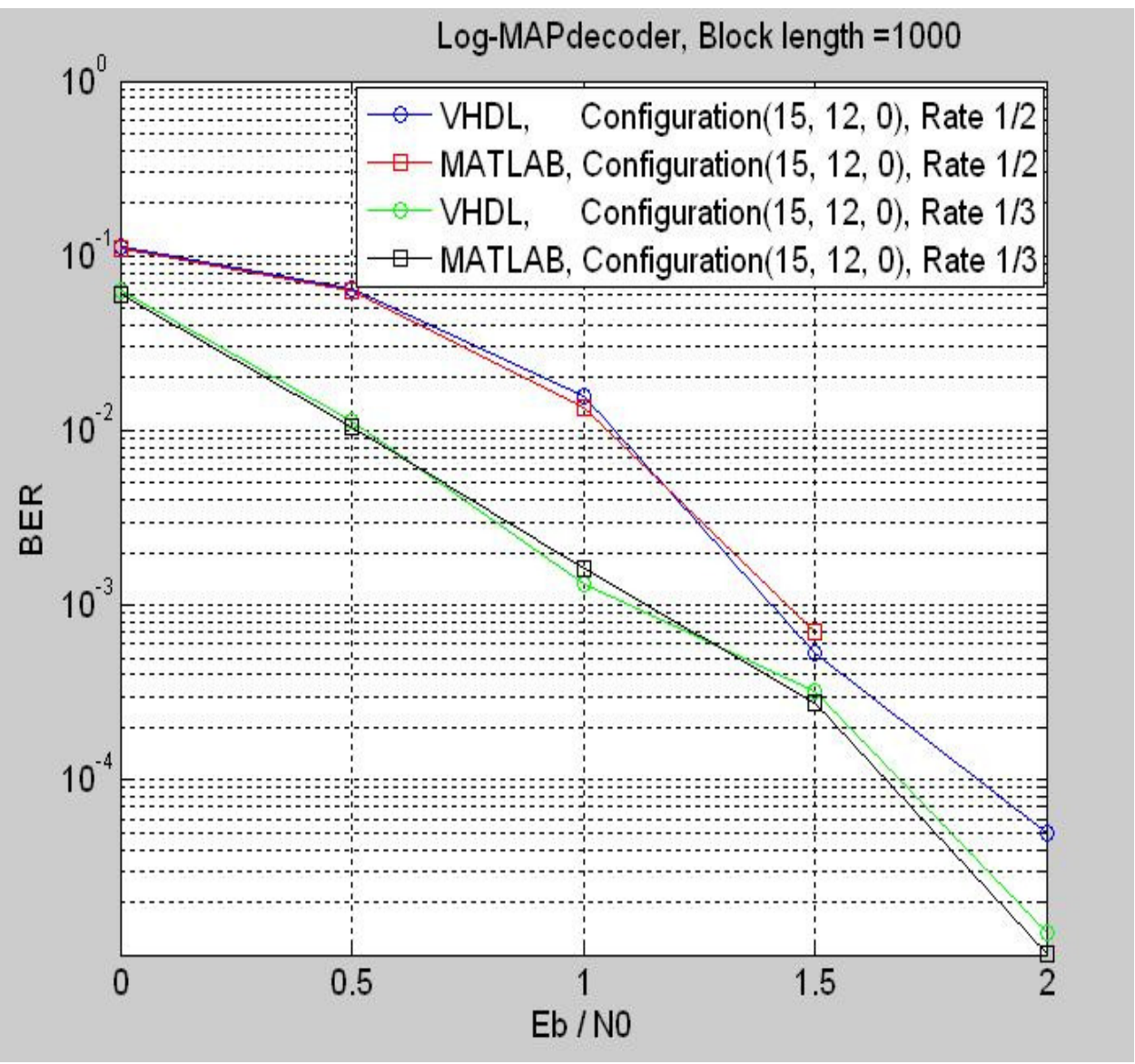

Fig. 6: Comparison of performance (between data obtained Using VHDL and MATLAB), for Rate 1/2 and Rate 1/3

From Fig. 6, it is clear that for lower SNR (Eb / N0), value performance obtained using VHDL encoded output data is comparatively similar with performance obtained using MATLAB encoded output data.

If we go for higher value of SNR then it show the variation between VHDL and MATLAB simulated results.

The performance in case of VHDL is marginally better than MATLAB. The above given Fig. 6 corresponds to rate 1/2 and 1/3, iteration 6 and 1000 block size data.

Now, we can go for higher number of memory element structute. Some of the following results are given below. These all results are corresponding to VHDL simulation.

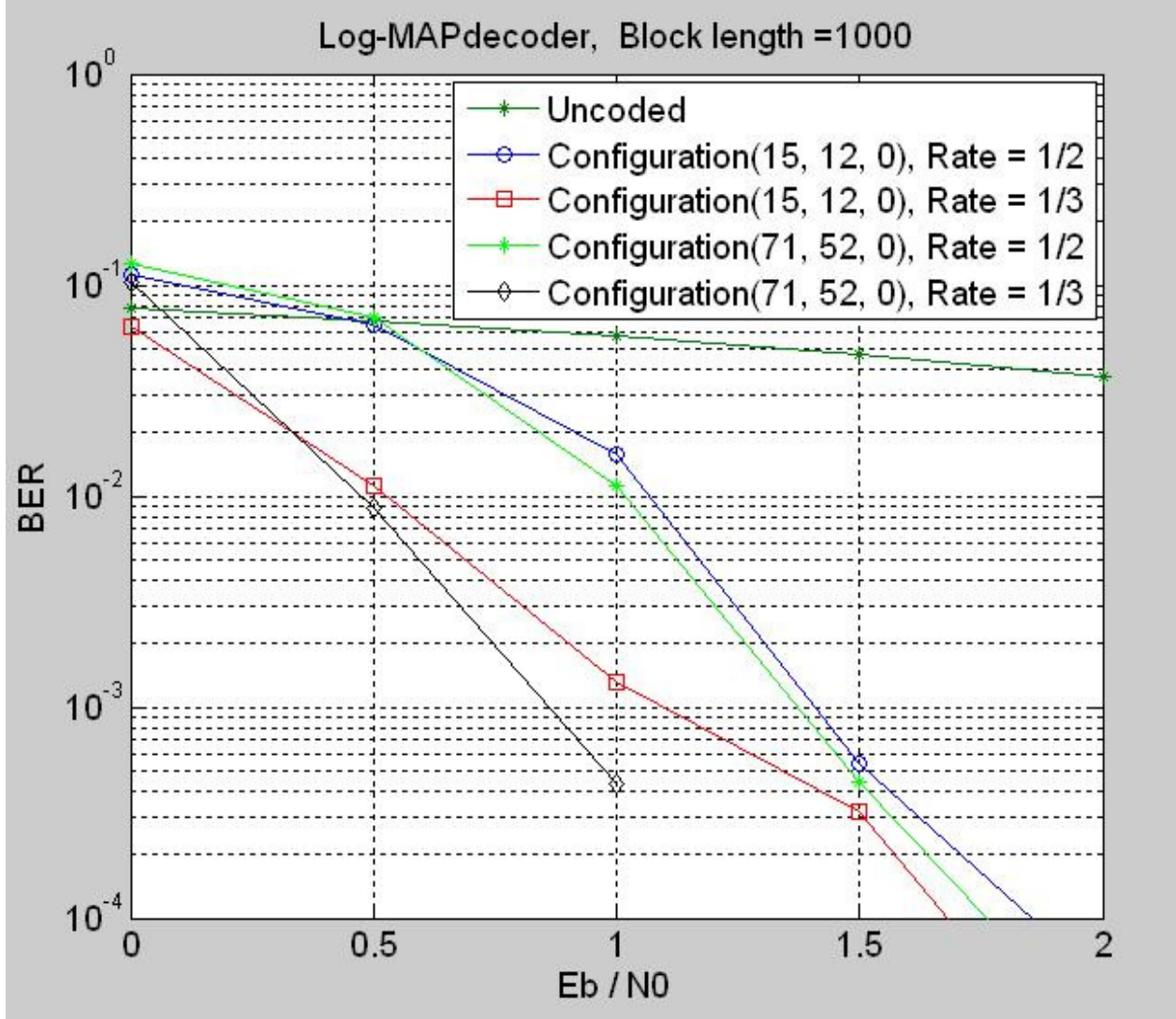

Fig. 7: Rate 1/2 vs. Rate 1/3 for different configuration (6 iteration)

From the above graph it is clear that if we go from rate 1/2 to rate 1/3, for any configuration of encoder, performance will improve. Here, we checked for two configurations of encoder, first (15, 12, 0) and second (71, 52, 0). For both of the configuration performance is satisfied.

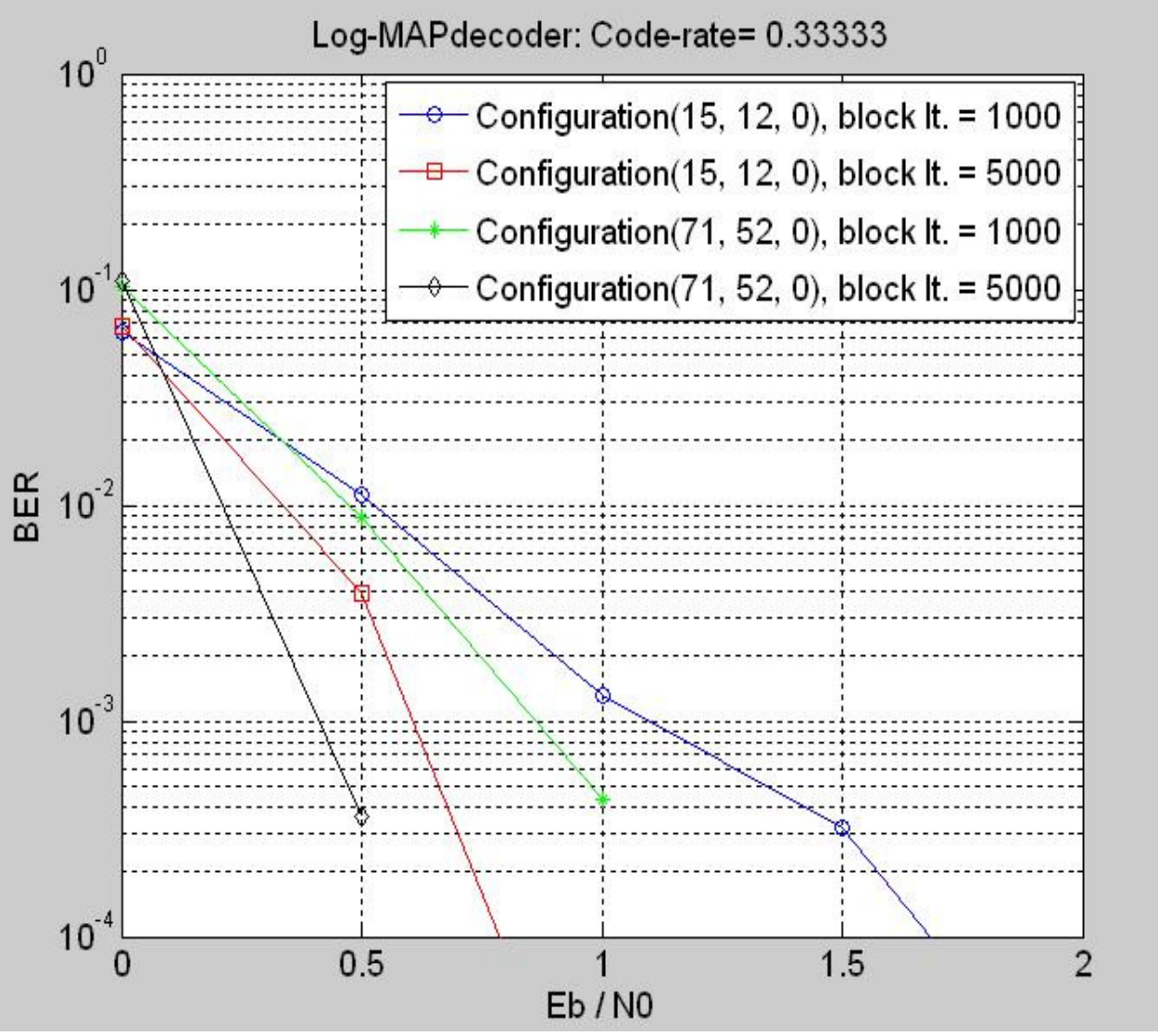

Fig. 8: performance comparison due to change in Block length (iteration 6)

Here, we have taken result corresponding to iteration 6, because we can see in Fig. 10, it is clearly mention that if we go for 1 iteration, performance is very poor as compare to further improved iteration.

When we go up to 5 or 6 iteration there are huge performance variation exist, but for further 7 or 8 iteration performance improves marginally. Whereas for next iterations, it not shows any improvement. So, for the purpose of reducing time delay, we have chosen iteration 6.

From the given Fig. 8, it shows that if we change the block length size then the performance will marginally improve. The graph given in Fig. 8, show comparison for given two configurations according to block length 1000 & 5000 and these all are corresponding to rate 1/3 and iteration 6.

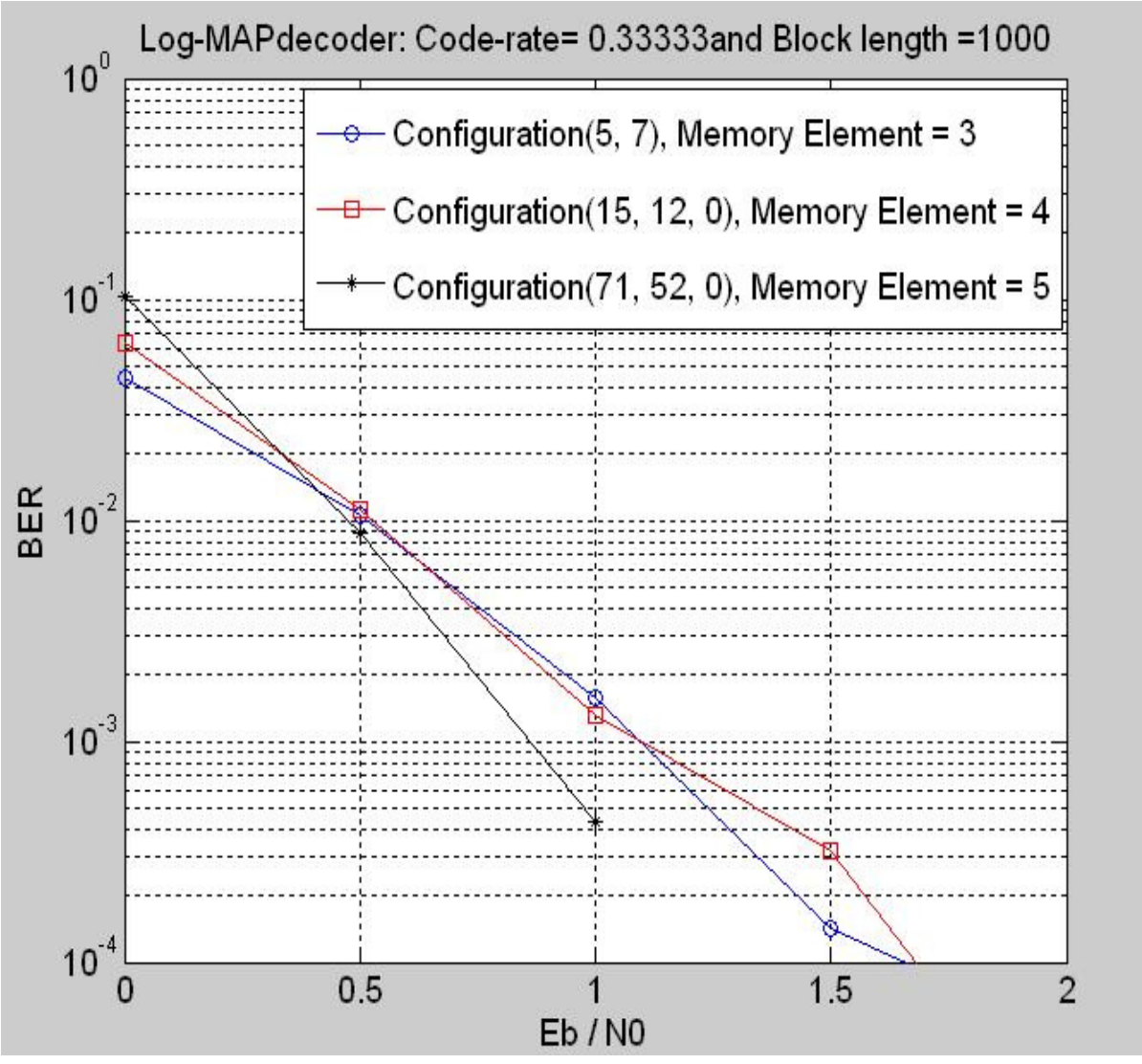

Fig. 9: performance comparison due change in Memory element (iteration 6)

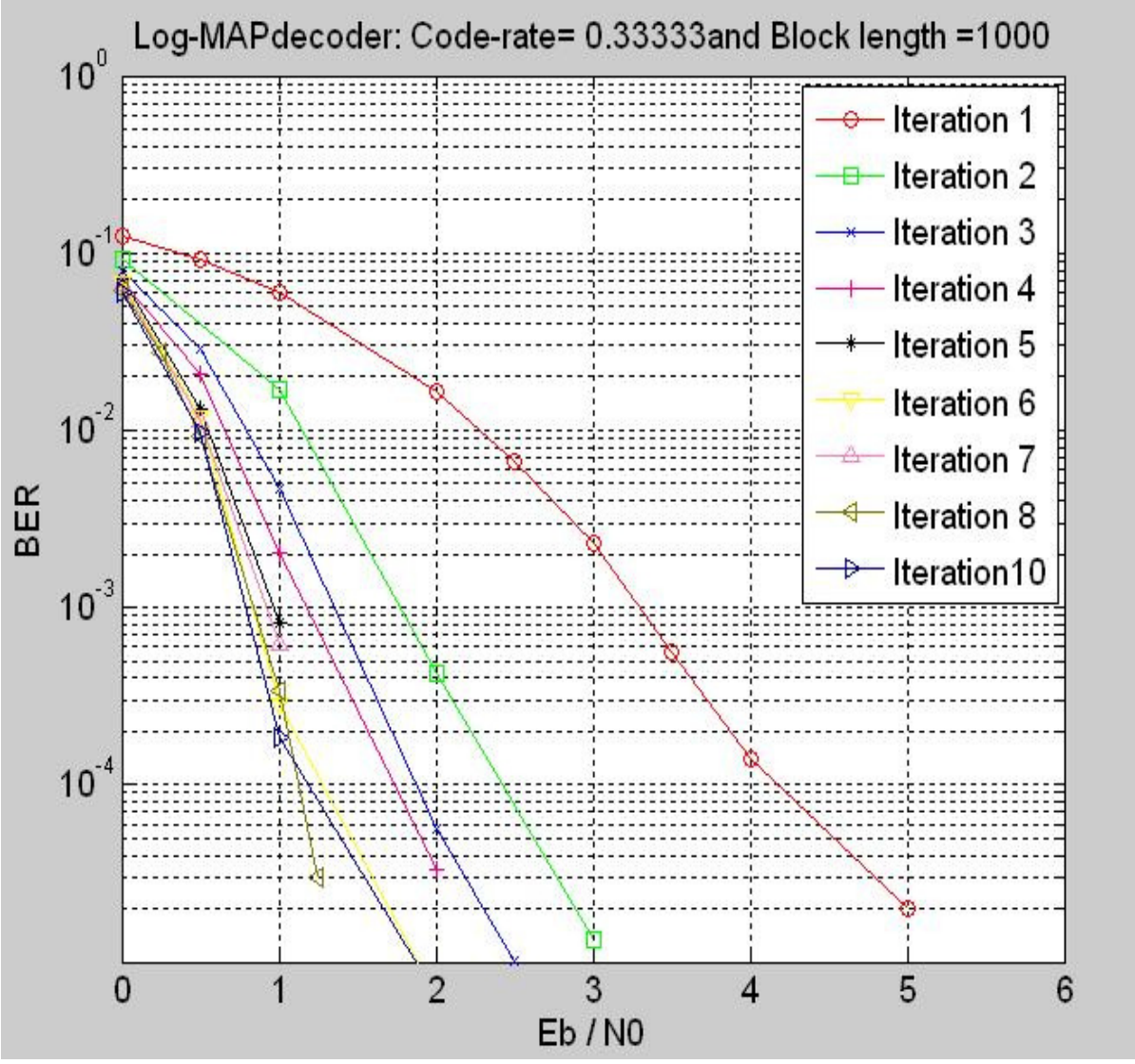

Fig. 10: Performance of turbo coding using Log MAP algorithm

We have seen performance comparison due to change in number of memory elements in Fig. 9.

From the above Fig. 9, it is clear that if choose higher number of memory elements in our encoder structure, performance improves. In this Fig. we have compared three different configuration of encoder using Log-MAP decoder, according to rate 1/3, block length 1000 and corresponding to iteration 6.

From the above Fig. 10, we have plotted the curve up to iteration 10 and observed that when we go from iteration 8 to 10, performance improvement not shown and curves overlap each other. This is reason to choose iteration 6, with this iteration time delay is also less.

## 4 CONCLUSION

In this paper, a design of different turbo encoder configuration like (15, 12, 0), (71, 52, 0) etc. are proposed through VHDL. These configurations give better performance and also show the lower computational complexity. Configuration (71, 52, 0) gives much better performance with less computational complexity. We can go for higher constraint length in VHDL for better BER performance. Turbo code is designed using Log-MAP decoder, and the obtained results are compared and verified with MATLAB simulated results of turbo code. We also compared the result obtained from VHDL simulation, with MATLAB simulated result and the reported result, for configuration (7, 5).

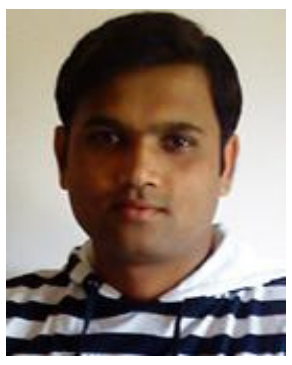

**Mr. Akash Kumar Gupta** has received his B.E. in 2008 from R.G.P.V. University, Bhopal, India. Presently he is pursuing M.E. from Birla Institute of Technology, Mesra (Jharkhand)-India, in the field of Electronics and Communication Engineering.

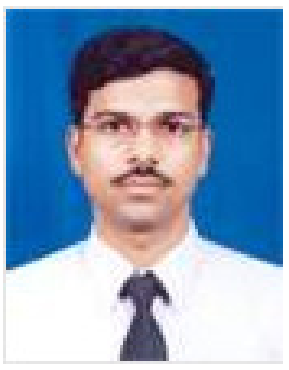

**Mr. Sanjeet Kumar** has received his B.sc (Engg.) from Magadh University, India, in 2001 and ME from BIT, Mesra, Ranchi, India, 2002 both in Electronics and Communication Engineering. Currently he is working as a Senior Lecturer in the Dept. of ECE, BIT, Mesra, Ranchi, India, and His Research interests are in area of Wireless video communication, information and coding theory.